\documentclass[a4paper,twocolumn, pre,showpacs,footinbib]{revtex4}

\usepackage{amsmath,amssymb}
\usepackage{makeidx}
\usepackage{amsfonts}
\usepackage[ansinew]{inputenc}
\usepackage[usenames,dvipsnames]{pstricks}
\usepackage{epsfig,graphicx}

\newcommand{\rme}{\mathrm{e}}
\newcommand{\rmi}{\mathrm{i}}
\newcommand{\rmd}{\mathrm{d}}
\newcommand{\bs}{\mathbf{s}}
\newcommand{\bsigma}{\mathbf{\sigma}}
\newcommand{\btau}{\mathbf{\tau}}
\newcommand{\Prob}{\mathrm{P}}
\newcommand{\W}{\mathrm{W}}
\newcommand{\M}{\mathrm{M}}

\makeindex

\begin{document}

\title{Islands of equilibrium in a dynamical world}

\author{David Saad}
\author{Alexander Mozeika}

\affiliation{The Non-linearity and Complexity  Research Group, Aston University, Birmingham B4 7ET, UK.}

\date{\today}

\begin{abstract}
Many natural, technological and social systems are inherently not in equilibrium. We show, by detailed analysis of exemplar models, the emergence of equilibrium-like behavior in localized or non-localized domains within non-equilibrium systems as conjectured in some real systems. Equilibrium domains are shown to emerge either abruptly or gradually depending on the system parameters and disappear, becoming indistinguishable from the remainder of the system for other parameter values. The models studied, defined on densely and sparsely connected networks, provide a useful representation of many real systems.
\end{abstract}



\maketitle

Equilibrium is a fundamental concept in statistical physics~\cite{LandauLifshitz1969}; it assumes that while the system dynamics is governed by microscopic interactions, some systems eventually reach a state where macroscopic observables remain unchanged. The evolution of such systems is driven by the corresponding Hamiltonian energy function and their states converge to the equilibrium distribution which is a function of energy only; all macroscopic properties of the system then follow from this distribution.

The dynamics of a non-equilibrium system, on the other hand, is typically not governed by a process derived from a Hamiltonian and such systems do not converge to an equilibrium state~\cite{CoolenSherrington93,DissipDynSys}. This is assumed to be true for many real systems, for instance in the financial, social and biological areas. However, constituents of some of these systems exhibit equilibrium-like behavior in emerging localized or non-localized domains; notable examples of this behavior are the emergence of equilibrium-like structures in functional brain networks~\cite{Fraiman}, neuronal dynamics~\cite{Gerhard} and the theory of markets~\cite{Foley}. Consequently, such domains may exist under some conditions within many other non-equilibrium systems but are difficult to identify.

Most systems in statistical physics fall into one of these two categories~\cite{Balescu}; the evolution of both equilibrium and non-equilibrium systems (in discrete time steps) is characterized by a trajectory $\bs(0)\rightarrow\cdots\rightarrow\bs(t)$, where $\bs(t)$ is a microscopic state of the system (\emph{microstate}) at time $t$. For Markovian processes this probability can be decomposed to a chain of transition probabilities from one time step to the next resulting in the joint probability
\begin{eqnarray}
\mathrm{P}[\bs(0)&\rightarrow&\cdots\rightarrow\bs(t)]= \label{def:ProbOfPath}\\ &\phantom{=}&\mathrm{W}[\bs(t)\vert\bs(t-1)]\times\cdots\times \mathrm{W}[\bs(1)\vert\bs(0)]\;\mathrm{P}(\bs(0))~,\nonumber
\end{eqnarray}
with initial $\mathrm{P}(\bs(0))$ and transition $\mathrm{W}[\bs(t)\vert\bs(t-1)]$ probability distributions. Expectation value of any macroscopic observable $M(\bs(t))$, i.e., a function of microstates defining a \emph{macrostate}, can be computed from the probability distribution (\ref{def:ProbOfPath}). Unfortunately, even for highly stylized models of statistical physics this procedure is  non-trivial~\cite{MimuraAndCoolen}. In equilibrium systems, one assumes that the probability of any microscopic trajectory is invariant under time-reversal; this leads to a property termed \emph{detailed balance} for the stationary distribution $\mathrm{P}_{\infty}(\bs)$ of process (\ref{def:ProbOfPath}), where transitions from state $\bs$ to $\hat{\bs}$ are balanced by transitions in the opposite direction $\mathrm{W}[\hat{\bs}\vert\bs]\;\mathrm{P}_{\infty}(\bs)=\mathrm{W}[\bs\vert\hat{\bs}]\;\mathrm{P}_{\infty}(\hat{\bs})$. For thermodynamic systems, this gives rise to the Gibbs-Boltzmann distribution $\mathrm{P}_{\infty}(\bs)\propto\mathrm{e}^{-\frac{1}{k_{B}T}E(\bs)}$, with temperature $T$, Boltzmann constant $k_B$ (we set $k_B=1$ for convenience)  and Hamiltonian (or energy) function $E(\bs)$, which usually follows from the transition probability $\mathrm{W}[\bs\vert\hat{\bs}]$~\cite{Peretto}. The stationary distributions in systems without detailed balance (when such distributions do exist) are generally much more complicated and difficult to analyze~\cite{CoolenSherrington93, DissipDynSys}.

In the absence of explicit time dependence, equilibrium systems therefore admit a reduced representation with respect to non-equilibrium ones, via the macrostates of the relevant (energy) functions. Some non-equilibrium physical systems show a local equilibrium-like behavior (e.g., having a slowly changing temperature) that allows for a similar reduced representation~\cite{Balescu}; however, this requires full knowledge of the corresponding Hamiltonian, which is completely unknown in many systems, especially in biological, financial and technological systems.

In past studies equilibrium and non-equilibrium systems analyses were typically well separated. In this Letter we show that in a large class of non-equilibrium systems, without detailed balance, one can still find domains that exhibit \emph{equilibrium-like}~\footnote{The term equilibrium-like behavior refers to systems where the average values of macroscopic observables are equal to those of their equilibrium counterparts, assuming that in very large systems only a limited number of observables can be measured.} behavior; these may be of a non-localised nature and may emerge and disappear depending on external conditions. In order to demonstrate this we study two exemplar models where one may intuitively anticipate this type of behavior to occur, and equally importantly, can quantitatively analyse it.

The two models considered here are Ising-like systems comprising $N$ spins $s_i\in\{-1,1\},~i\in\{1,\ldots,N\}$, representing variables (degrees of freedom) interacting on sparsely and densely connected networks. This type of system is commonly used in statistical physics as a prototype and a first approximation in modelling complex phenomena in many-body systems~\cite{Dorogovtsev}.
In the densely connected model each variable interacts with a very large (order of the system size) number of variables whereas in the sparse model the number of interactions is much smaller than that of the system size. Furthermore, both models have  bipartite topologies where one part of the network serves as a non-equilibrium ``environment'' while the other is designed to be in equilibrium {\em when considered on its own}.

\begin{figure*}
\setlength{\unitlength}{1mm}
\begin{picture}(200,100)
\put(-5,55){\epsfysize=70\unitlength\rotatebox{0}{\epsfbox{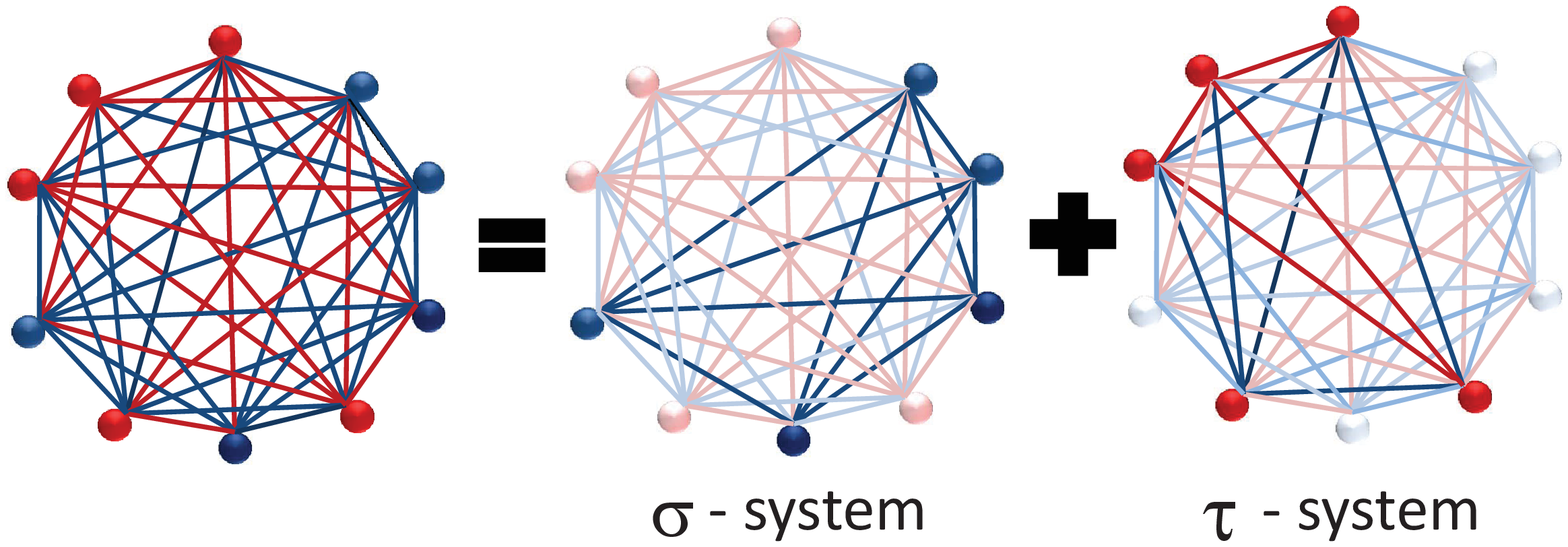}}}\put(5,100){$(a)$}
\put(-5,0){\epsfysize=70\unitlength\rotatebox{0}{\epsfbox{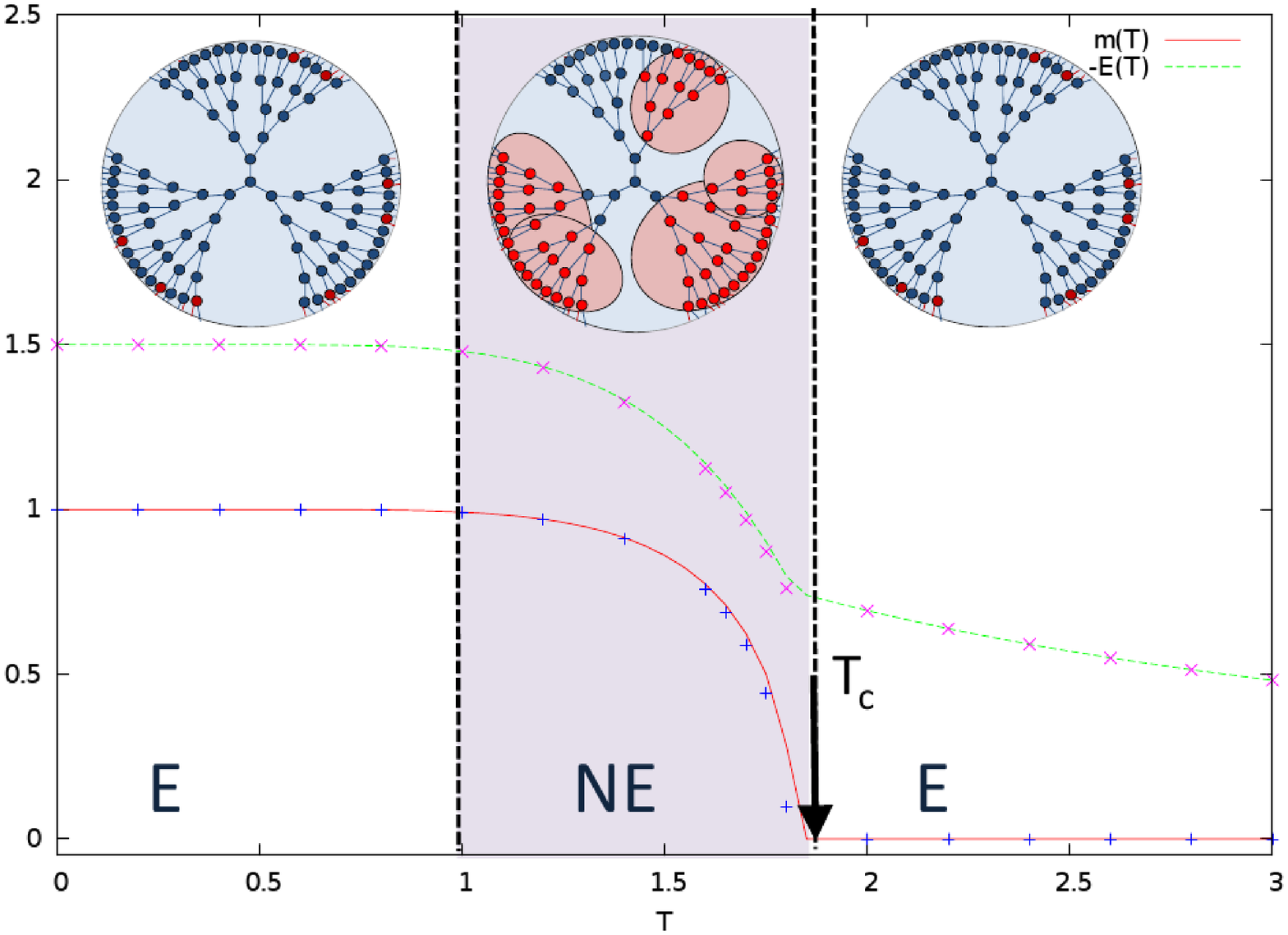}}}\put(5,65){$(b)$}
%
\put(90,57){\epsfysize=53\unitlength\epsfbox{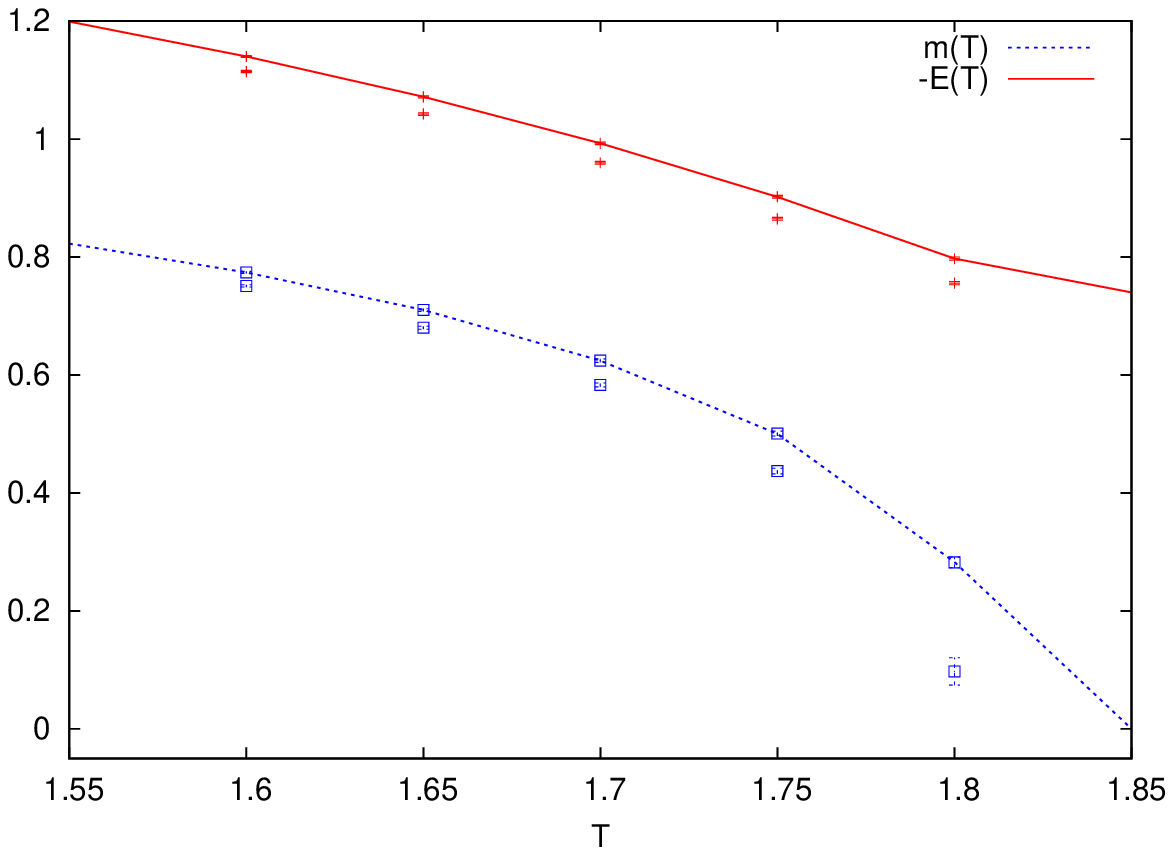}}\put(100,100){$(c)$}
\put(88,2){\epsfysize=54.5\unitlength\epsfbox{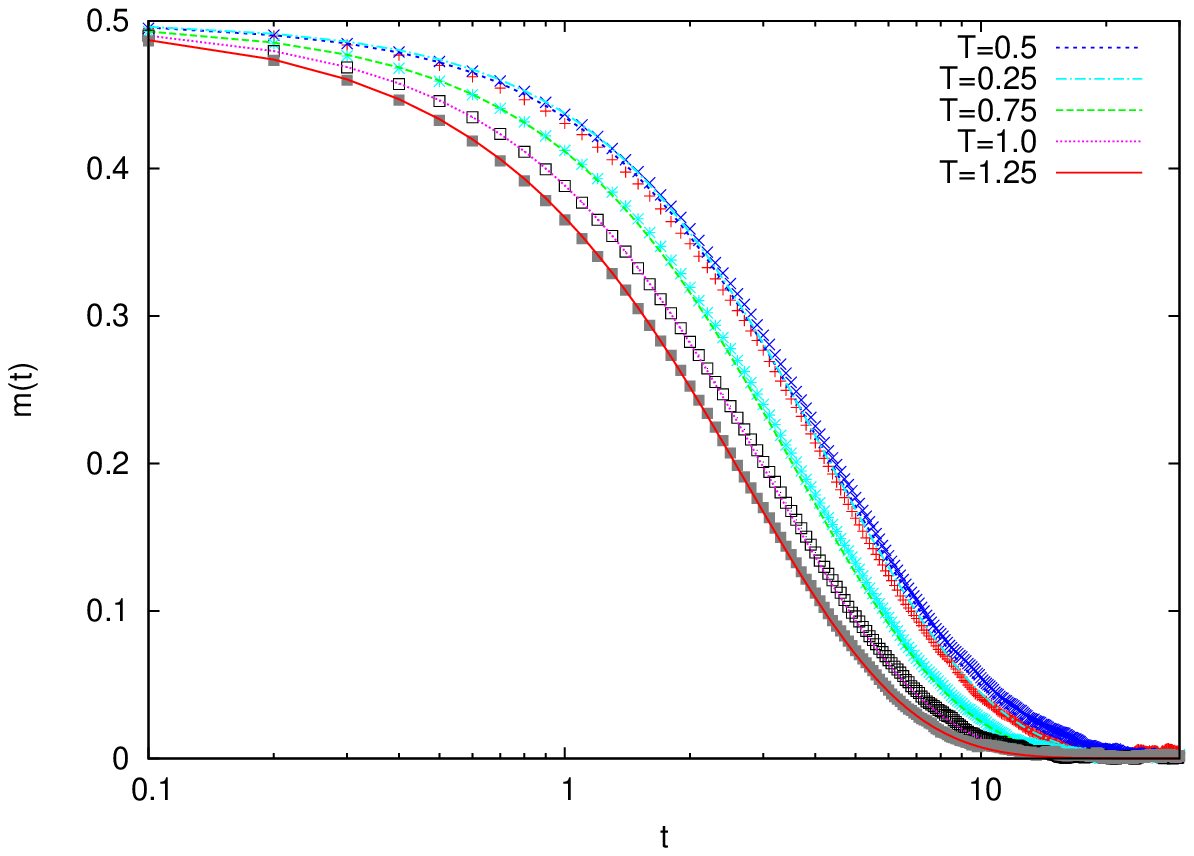}}\put(100,45){$(d)$}

\end{picture}
%
\caption{(a) Densely connected system composed of equilibrium ($\sigma$-system, blue nodes) and non-equilibrium ($\tau$-system, red nodes) components. Blue and red edges represent positive and negative interactions, respectively. Interaction directions are not shown.  (b) Properties of sparse systems exhibiting equilibrium-like behavior.  The degrees of freedom (blue nodes) are interacting on graphs with locally  tree-like topology.  A fraction of these nodes (red) are exposed to the changing environment (perturbations).  Macroscopic observables of the un-perturbed nodes suggest they are in thermal equilibrium at low temperature as the influence of perturbations on the macroscopic observables is negligible (left - E panel); as one approaches a critical temperature, the system becomes very sensitive, develops long-range order and exhibits significant deviations from the equilibrium values of these observables (middle - NE panel). The transition point is determined by the point where deviation from the equilibrium values exceed thermal fluctuations and represents an estimate. The perturbations become negligible again at the high temperature region as one moves away from the critical temperature $T_c$ (right - E panel). This qualitative explanation is supported by comparing the equilibrium energy $E$ (dashed line) and magnetization $m$ (solid line) with the average energy  $E(\sigma)=-\frac{1}{N}\sum_{\langle ij\rangle}\sigma_i
\sigma_j$ and magnetization densities $m(\sigma)=\frac{1}{N}\sum_{i=1}^N\sigma_i$ measured in Monte Carlo simulations (symbols) of a ferromagnetic Ising spin system, defined on a random regular graph of size $N=10^6$ with $k=3$, where a fraction $p=0.05$ of sites are subject to the external time-dependent  random binary fields  $\theta_i(t)\in\{-1,1\}$, with $\Prob(\theta_i(t)=\pm1)=1/2$; all Monte Carlo simulations results reported here have been carried out for a similar system size and connectivity degree. The measurements are taken only on sites not influenced by $\theta_i(t)$. 
(c) Deviations from equilibrium, shown in NE panel of (b), are much larger than one usually finds due thermal fluctuations alone in equilibrium. We compare thermal fluctuations of the equilibrium energy $E$ (dashed line) and magnetization $m$ (solid line) with those measured in Monte Carlo simulations (represented by symbols with error bars, much smaller than the symbol size, on the solid lines). The non-equilibrium simulation measurements (symbols with error bars) are taken only at sites not influenced by $\theta_i(t)$ and show clear deviation from the equilibrium values. (d) Comparing the magnetization $m$ (solid lines) calculated theoretically with values (symbols) measured in the Monte Carlo simulations of ferromagnetic Ising spin system defined on an asymmetric (an incoming edge with probability $1/2$) random regular graph show good agreement between the two.
\label{fig:1} }
\end{figure*}
\emph{Densely connected model:} This model, described schematically in Figure~\ref{fig:1}(a), is governed by the  process (\ref{def:ProbOfPath}) where the microstate  $\bs(t)=(\sigma_1(t),\ldots,\sigma_{N^{\sigma}}(t),\tau_1(t),\ldots,\tau_{N^\tau}(t))$ is represented for clarity by two components consisting of $N^\sigma$ and $N^\tau$ extensive degrees of freedom, respectively, such that $N^\sigma+N^\tau=N$; the distinction between the two subsystems is not obvious through interaction strengths. The $\tau$-component of the system ($\tau$-system) drives the $\sigma$-component ($\sigma$-system) via stochastic alignment of spins $\{\sigma_i\}$ to the corresponding local fields~$h_i(\bsigma,\btau)=\sum_{j\neq i}^{N^\sigma}J_{ij}^{\sigma}\sigma_j+\sum_{j}^{N^\tau}J_{ij}^{\sigma\tau}\tau_j+\theta_i^\sigma $ and is itself governed by the stochastic alignment of $\{\tau_i\}$ to the local fields  $g_i(\bsigma,\btau)=\sum_{j}^{N^\sigma}J_{ij}^{\tau\sigma}\sigma_j+\sum_{j\neq i}^{N^\tau}J_{ij}^{\tau}\tau_j+\theta_i^\tau$, where  the variables $\{J_{ij}^{\sigma}, J_{ij}^{\tau},  J_{ij}^{\sigma\tau},J_{ij}^{\tau\sigma}\}$ prescribe the strengths of the various interactions  and $\{ \theta_i^\sigma,\theta_i^\tau\}$ are external fields which may depend on time. Each site in the $\sigma$-system ($\tau$-system)  is updated in a stochastic manner with the probabilities $\Prob[\sigma_i(t+1)]\propto\exp[\beta\sigma_i(t+1) h_i(\bsigma(t),\btau(t))]$ and $\Prob[\tau_i(t+1)]\propto\exp[\beta\tau_i(t+1) g_i(\bsigma(t),\btau(t))]$, respectively, which are controlled by the noise parameter $\beta$  (that defines the temperature $T=1/\beta$); the dynamics is completely deterministic when $\beta\rightarrow\infty$ and is completely random when  $\beta=0$. All sites are updated independently of each other, which leads to the Markov process (\ref{def:ProbOfPath}).

It is clear from the definitions of the fields that the two systems evolve independently and separately when all cross-component interactions $J_{ij}^{\sigma\tau}=J_{ij}^{\tau\sigma}=0$. If in addition all $J_{ij}^{\sigma}$ are symmetric, i.e. $J_{ij}^{\sigma}=J_{ji}^{\sigma}$, and all external fields $\theta_i^\sigma(t)$ do not depend on time, then the $\sigma$-system is governed by the equilibrium distribution $\mathrm{P}_{\infty}(\bsigma)\propto\mathrm{e}^{-\beta E_\beta(\bsigma)}$ with Peretto's~\cite{Peretto} pseudo-Hamiltonian
 \begin{equation}
E_\beta(\bsigma)=-\frac{1}{\beta}\sum_{i=1}^{N^\sigma}\log2\cosh[\beta h_i(\bsigma,0)]-\sum_{i=1}^{N^\sigma}\theta_i^\sigma\sigma_i.\label{def:H}
\end{equation}
For asymmetric cross-component interactions $J_{ij}^{\sigma\tau}\neq0$ or $J_{ij}^{\tau\sigma}\neq0$ the \emph{complete} system  is not in equilibrium. However, \emph{this does not prevent the $\sigma$-system from exhibiting equilibrium-like behavior}. To see this we consider the simplest case of $J_{ij}^{\sigma}=J_{ij}^{\sigma\tau}=1$ and  $J_{ij}^{\tau}\neq J_{ji}^{\tau}$, where both interaction variables $J_{ij}^{\tau}$ and $J_{ij}^{\tau\sigma}$ are independent random variables and are assigned values of $\pm1$ with equal probability; to simplify the example we will also choose $N^\sigma=N^\tau=N/2$~\footnote{As we focus on the thermodynamic limit, a more careful scaling of the interactions is required. See Appendix for a detailed derivation that also accommodates more general interaction strengths and interaction disorder.}.
%

We employ the method of generating functional analysis to obtain expectation values of various macroscopic quantities, averaged over the quenched disordered induced by the randomly assigned values of $J_{ij}^{\tau}$ and $J_{ij}^{\tau\sigma}$.
It turns out that in this case the complete system admits a macroscopic description via the magnetizations $m^\sigma(\sigma(t))=\frac{1}{N^\sigma}\sum_{i=1}^{N^\sigma}\sigma_i(t)$ and $m^\tau(\tau(t))=\frac{1}{N^\tau}\sum_{i=1}^{N^\tau}\tau_i(t)$. In particular, for the magnetizations averaged over the process, $m^\sigma(t)=\left\langle m^\sigma(\sigma(t))\right\rangle$, $m^\tau(t)=\left\langle m^\tau(\tau(t))\right\rangle$, and in the thermodynamic limit $N\rightarrow\infty$, one obtains
\begin{eqnarray}
&&m^\sigma(t+1)=\tanh\beta[m^\sigma(t) +  m^\tau(t)   +\theta^\sigma(t)]\label{eq:m}\\
&&m^\tau(t+1)=0\nonumber
\end{eqnarray}
with initial conditions given by $m^\sigma(0)$ and $m^\tau(0)$. For $\theta^\sigma(t)=\theta^\sigma$ and $\theta^\tau =0$ this equation admits a stationary solution~$m^\sigma(\infty)=\tanh[\beta( m^\sigma(\infty)  +\theta^\sigma                )]$ which is exactly the same as one finds in equilibrium~\cite{TheorOfIP} governed by (\ref{def:H}). Similar argument also holds for the average density $-\frac{1}{N^\sigma}\frac{1}{\beta}\sum_{i=1}^{N^\sigma}\log2\cosh[\beta h_i(\sigma,\tau)]-\theta^\sigma \frac{1}{N^\sigma}\sum_{i=1}^{N^\sigma}\sigma_i$, which approaches the equilibrium energy (\ref{def:H}) and is a function of the magnetization only. Furthermore, for $\theta^\sigma=0$ the stationary solution $m^\sigma(\infty)=0$ (disordered phase) is stable when $\beta <1$ but  bifurcates into two solutions $\vert m^\sigma(\infty)\vert\neq0$ (ordered phase) at $\beta =1$.  Thus both parts of the system are indistinguishable when $\beta <1$.

While we deliberately focussed on a particularly simple and tractable model, more complex systems of similar characteristics could be constructed to demonstrate the existence of equilibrium-like domains in a non-equilibrium environment.

\emph{Sparsely connected model:}
The model considered here is a sparsely connected Ising ferromagnetic system defined on an $N$-node random regular graph where each node is  randomly connected to exactly $k\!\in\! O(N^0)$ other nodes. The system evolves by selecting a node $i$ with probability $1/N$ at each time step and aligning its state $\sigma_i$ to the local field $h_i(\sigma)\!=\!J\sum_{j\in \partial i} \sigma_j$ with  probability proportional to $\rme^{\beta\sigma_i h_i(\sigma)}$ , where $\partial i$  is a set ($\vert\partial i\vert\!=\!k$) of sites directly connected to site $i$. This leads to a Markovian process in continuous time (see Appendix). Furthermore, a  fraction $p$ of (randomly selected) spins  in this system  are driven by the random time-dependent external fields $\theta_i(t)\!\in\!\{-1,1\}$, where $\Prob(\theta_i(t)\!=\!\pm1)\!=\!1/2$, i.e., in these sites the field $h_i(\sigma)$ is effectively changed to  $h_i(\sigma)\!+\!\theta_i(t)$.

\emph{Without}  external fields and after long time ($t\!\rightarrow\!\infty$) the system is in thermal equilibrium and the spins are governed by the Gibbs-Boltzmann distribution with the Hamiltonian $E(\sigma)\!=\!-J\sum_{\langle ij\rangle}\sigma_i\sigma_j$. In the equilibrium  the average energy and magnetization are given respectively by the equations
\begin{eqnarray*} E&=&-\frac{1}{2}k\frac{\tanh(\beta J)+\tanh(\beta h)^2}{1+\tanh(\beta J)\tanh(\beta h)^2} \\ m&=&\tanh\{\tanh^{-1}[\tanh(\beta J)\tanh(\beta h)]k\}~,\end{eqnarray*}
respectively,
where $h$ is a solution of $h=\frac{1}{\beta}(k\!-\!1)\tanh^{-1}[\tanh(\beta J)\tanh(\beta h)]$~\cite{BetheSG}. The system is in an ordered (disordered) state if $T\!<\!T_c$ ($T\!>\!T_c$), with $T_c\!=\!J/\tanh^{-1}\frac{1}{k-1}$ being the critical temperature of the system. In the presence of time-dependent external fields convergence to thermal equilibrium  is no longer guaranteed, but part of the system, which is not directly affected by the external fields, can exhibit equilibrium-like behavior as can be seen in Figure~\ref{fig:1}(b). This phenomena, vanishes when the temperature $T$ in the system is close to $T_c$. The presence of this phase transition seems to magnify the non-equilibrium effect of an external driving field which is much larger than one usually finds due to the thermal fluctuations  alone, in equilibrium, as can be seen in Figure~\ref{fig:1}(c). We note that similar behavior also occurs in a system defined on a Cayley tree where boundary sites are subject to the same external fields~\cite{SM}.

Alternatively,  the $\theta_i$ can be viewed as a field induced by a non-equilibrium part of the system.  In the long time limit $t\rightarrow\infty$,  this system is equivalent to the setup where one part of the system (asymmetric) drives the other (symmetric). The sites affected by the asymmetric part are described by the set $\{m_i(t)\}$ of local magnetizations $m_i(t)=\sum_\sigma \Prob_t(\sigma)\;\sigma_i$. Furthermore, if the stationary point of these local magnetizations is exactly $m_i(t)=0$, the asymptotic behavior of the system is equivalent to that of the system depicted in  Figure~\ref{fig:1}(b).

To verify this we assume that for asynchronous dynamics on an asymmetric regular graph the local magnetization $m_i(t)$ is a function of the local magnetizations $m_j(t)$ of its neighbors $j\in\partial i$ only. For $k=3$ this leads to the following set of equations
\begin{eqnarray}
\frac{\rmd}{\rmd t}m_i+m_i= \left\{
\begin{array}{l l}
  (A+7\Gamma)\sum_{j\in\partial i} m_j & \quad \mbox{if $\vert\partial i\vert=3$}\\
   ~~~~~+6\Gamma\prod_{j\in\partial i} m_j & \label{eq:m_i}\\
  \frac{1}{2}\tanh(2\beta)\sum_{j\in\partial i} m_j & \quad \mbox{if $\vert\partial i\vert=2$}\nonumber\\
  \tanh(\beta)\sum_{j\in\partial i} m_j& \quad \mbox{if $\vert\partial i\vert=1$}\\
  0& \quad \mbox{if $\vert\partial i\vert=0$}\\
\end{array} \right.
\end{eqnarray}
where $A\!=\!(27\tanh(\beta)\!-\!\tanh(3\beta))/24$ and $\Gamma\!=\!(\tanh(3\beta)\!-\!3\tanh(\beta))/24$, which is valid for single instances of asymmetric regular graphs as can be seen in Figure~\ref{fig:1}(d).

Recent studies of neural populations~\cite{Schneidman}, flocks of birds~\cite{Bialek}, magnets~\cite{Robb} and of many other natural and technological systems, suggest the existence of equilibrium domains in non-equilibrium systems. However, to show the emergence of such domains in practice may prove difficult, especially if they are composed of \emph{non-localized} degrees of freedom; for instance, a group of traders located in different stock markets and aiming
to maximize their profits may (possibly inadvertently) constitute an
equilibrium-like system. This Letter aims to change our viewpoint on the traditional separation between equilibrium and non-equilibrium systems in order to understand the emergence of equilibrium behaviors within non-equilibrium systems and possibly facilitate control of this phenomenon. The exemplar models systematically analyzed here represent the first step towards this goal; they demonstrate the emergence of such domains and their dependence on various system parameters as well as their dissipation close to criticality. In the real world such systems may emerge randomly or evolve in a structured manner through a selection process. The study opens up exciting opportunities for future work on the role and dynamics of equilibrium domains in systems with adiabatically changing interactions and parameters, such as coordinated global trade and social networks.

 \begin{acknowledgments}We would like to thank David Sherrington, Marc M\'ezard, David Lowe and Riccardo Zecchina for very helpful comments on the manuscript. This work is supported by the EU FET project STAMINA (FP7-265496)  and the Leverhulme trust grant F/00 250/H. \end{acknowledgments}

\appendix

\section{Processes on graphs}

We  consider a system of $N$ Ising spins, $\sigma_i\in\{-1,1\}$, which are placed on the vertices of a graph and interact only when they are connected.
Their microscopic dynamics are governed by a Glauber type stochastic algorithm where a spin on site $i$ is  flipped with probability
\begin{eqnarray}
\Prob(\sigma_i\rightarrow -\sigma_i)=\frac{\rme^{-\beta\sigma_i h_i(\sigma)}}{2\cosh(\beta h_i(\sigma))},\label{eq:algorithm}
\end{eqnarray}
where $h_i(\sigma)$ is
a local field defined as
\begin{eqnarray}
h_i(\sigma)=\sum_{j\in\partial i} J_{ij}\sigma_j+
\theta_i,\label{eq:field}
\end{eqnarray}
with $\partial i$ being the set of sites connected to site $i$ and where we have used the notation $\sigma=(\sigma_1,\ldots,\sigma_N)$. The parameter
$\beta$ controls the level of noise in the system; the
dynamics is completely random when $\beta\rightarrow0$ and completely deterministic when
$\beta\rightarrow\infty$. The parameter $\theta_i$ defines an
external field. The set of variables $\{J_{ij}\}$ prescribes the strengths of interactions between the spins. Once chosen these variables are kept fixed for the duration of the process.

In order to complete the above algorithm we have to specify how we choose the sites for each update according to (\ref{eq:algorithm}). A simplest choice is to update all sites simultaneously which gives rise to the parallel dynamics governed  by the Markov equation (this type of dynamics is popular in the modeling of neural networks~\cite{TheorOfIP})
\begin{eqnarray}
\Prob_{t+1}(\sigma) &=& \sum_{\sigma^\prime}W[\sigma\vert\sigma^\prime]\Prob_{t}(\sigma^\prime)\label{eq:MarkovParallel}
\end{eqnarray}
with  the transition probability
\begin{eqnarray}
\W[\sigma\vert\sigma^\prime]=\prod_{i=1}^N\frac{\rme^{\beta\sigma_i h_i(\sigma^\prime) }}{2\cosh(\beta h_i(\sigma^\prime))}.\label{eq:TransitionM}
\end{eqnarray}

For the symmetric interactions, i.e. $J_{ij}=J_{ji}$, the  detailed balance property $\W[\sigma^\prime\vert\sigma]\;\Prob(\sigma)=\W[\sigma\vert\sigma^\prime]\;\Prob(\sigma^\prime)$ is always satisfied. If in addition the ergodic property ($\exists\; t^\prime$ such that for $\forall\; t\geq t^\prime$: $\Prob_t(\sigma)>0$)  is  satisfied then the process (\ref{eq:MarkovParallel}) converges~\cite{Peretto} to the equilibrium distribution
\begin{equation}
\Prob_{\infty}(\sigma)\propto\mathrm{e}^{-\beta E_\beta(\sigma)},\label{eq:EqPar}
\end{equation}
where $E_\beta(\bsigma)$ is the pseudo-Hamiltonian (this is not a proper Hamiltonian because of its explicit dependence on the noise parameter $\beta$)
 \begin{equation}
E_\beta(\sigma)=-\frac{1}{\beta}\sum_{i=1}^{N}\log2\cosh(\beta h_i(\sigma))-\sum_{i=1}^{N}\theta_i\sigma_i.\label{def:Hperetto}
\end{equation}
A slightly more complicated scenario is when the sites of a system are updated asynchronously in the following manner: at each iteration of the  algorithm a site
$i$ is drawn randomly and independently from the set $\lbrace 1,\ldots, N\rbrace$ of
all sites then the spin $\sigma_i$ of this site is updated with the probability (\ref{eq:algorithm}) (this is one of the main algorithms used to study the dynamics of Ising-type magnetic systems~\cite{Nishimori}). This process naturally leads to the Markov
equation~\cite{Bedeaux} in continuous time
\begin{eqnarray}
\frac{\rmd}{\rmd t}\Prob_t(\sigma) &=& \sum_{i=1}^N \big [\Prob_t(\sigma_1,\!\ldots,\!-\!\sigma_i,\!\ldots\!,\sigma_N)\Prob(\sigma_i\!\rightarrow\!\sigma_i)\label{eq:master}\\
&&~~~~~~~~~~-\;\Prob_t(\sigma_1,\!\ldots\!,\sigma_N)\Prob(\sigma_i\!\rightarrow\! -\!\sigma_i)\big].\nonumber
\end{eqnarray}
As in the case of synchronous dynamics (\ref{eq:MarkovParallel}) the process (\ref{eq:master}) satisfies detailed balance only for symmetric interactions and it evolves towards the equilibrium Gibbs-Boltzmann
distribution $\Prob_\infty (\sigma)\propto\rme^{-\beta E(\sigma)}$,
with the Hamiltonian (or energy) function
\begin{eqnarray}
E(\sigma)=-\sum_{\langle ij\rangle}J_{ij}\sigma_i
\sigma_j-\sum_{i=1}^N \theta_i\sigma_i\label{def:Hg-b}
\end{eqnarray}
(the first sum is over all edges in the graph), which is a unique stationary solution when the process (\ref{eq:master}) is ergodic. 

\section{Dynamics of a densely connected model\label{section:dense}}

\subsection{Generating functional\label{section:GF}}
In this section we study dynamics of a densely connected Ising spin system governed by the Markov process  with the transition probability given by
\begin{widetext}
\begin{eqnarray} \mathrm{W}[\bsigma(t\!+\!1),\btau(t\!+\!1)\vert\bsigma(t),\btau(t)]&=&\prod_{i=1}^{N^\sigma}\frac{\rme^{\beta\sigma_i(t\!+\!1)h_i(\bsigma(t),\btau(t))}}{2\cosh[\beta h_i(\bsigma(t),\btau(t))]}\prod_{\ell=1}^{N^\tau}\frac{\rme^{\beta\tau_\ell(t\!+\!1)g_\ell(\bsigma(t),\btau(t))}}{2\cosh[\beta g_\ell(\bsigma(t),\btau(t))]}\label{def:TransProb},
\end{eqnarray}
\end{widetext}
where $ h_i(\bsigma,\btau)=\sum_{j\neq i}^{N^\sigma}J_{ij}^{\sigma}\sigma_j+\sum_{j}^{N^\tau}J_{ij}^{\sigma\tau}\tau_j+\theta_i^\sigma  $ and  $g_i(\bsigma,\btau)=\sum_{j}^{N^\sigma}J_{ij}^{\tau\sigma}\sigma_j+\sum_{j\neq i}^{N^\tau}J_{ij}^{\tau}\tau_j+\theta_i^\tau$. The averages of various macroscopic quantities in this system can be conveniently computed from the generating function
\begin{widetext}
\begin{equation}
\Gamma[\psi^\sigma,\psi^\tau]=\left\langle\exp\left[-\rmi\sum_{t=0}^{t_{\rm max}}\left\{\sum_{i=1}^{N^{\sigma}}\psi_i^\sigma(t)\sigma_i(t)+\sum_{\ell=1}^{N^{\tau}}\psi_\ell^\tau(t)\tau_\ell(t)\right\}\right] \right\rangle,\label{def:G}
\end{equation}
\end{widetext}
where the average $\left\langle \cdots\right\rangle$ is taken over the microscopic trajectories $\sigma(0),\tau(0)\rightarrow\cdots\rightarrow\sigma(t_{\rm max}),\tau(t_{\rm max})$ occurring with probability
\begin{equation}
\Prob(\sigma(0))\Prob(\tau(0))\prod_{t=0}^{t_{\rm max}-1}\mathrm{W}[\bsigma(t\!+\!1),\btau(t\!+\!1)\vert\bsigma(t),\btau(t)].
\end{equation}
Inserting into the generating function (\ref{def:G}) the following integral representations of $\delta$-functions for all times $t$ and site indices $i$
\begin{eqnarray}
\int\frac{\mathrm d  h_{i}(t)\;\mathrm d  \hat{h}_{i}(t)}{2\pi}\;\rme^{\rmi  \hat{h}_{i}(t)[ h_{i}(t)-
h_i(\bsigma(t),\btau(t))
]}\!=\!1\label{def:unity}\\
\int\frac{\mathrm d  g_{i}(t)\;\mathrm d  \hat{g}_{i}(t)}{2\pi}\;\rme^{\rmi  \hat{g}_{i}(t)[ g_{i}(t)-
g_i(\bsigma(t),\btau(t))
]}\!=\!1\nonumber
\end{eqnarray}
we obtain
\begin{widetext}
\begin{eqnarray}
\Gamma[\psi^\sigma,\psi^\tau]&=&\int \{\rmd h\; \rmd \hat{h}\; \rmd g\; \rmd \hat{g}\}
\exp\left[\rmi\sum_{t=0}^{t_{\rm max}\!-\!1}\left\{\hat{h}(t)\cdot h(t)+\hat{g}(t)\cdot g(t)\right\}\right]\label{eq:G1}\\
&&\times\sum_{\{\sigma_i(t),\tau_i(t)\}} \exp\left[-\rmi\sum_{t=0}^{t_{\rm max}\!-\!1}\left\{\hat{h}(t)\cdot h(\bsigma(t),\btau(t))+\hat{g}(t)\cdot g(\bsigma(t),\btau(t))\right\}\right]\nonumber\\
&&\times\Prob \big[\sigma(t_{\rm max}),\tau(t_{\rm max})\leftarrow\cdots\leftarrow\sigma(1),\tau(1)\nonumber\\
&&~~~~~~~~\vert h(t_{\rm max}\!-\!1),g(t_{\rm max}\!-\!1),\cdots,h(0),g(0)\big]\Prob(\sigma(0))\Prob(\tau(0))\nonumber\\
&&\times\exp\left[-\rmi\sum_{t=0}^{t_{\rm max}}\left\{\psi^\sigma(t)\cdot\sigma(t)+\psi^\tau(t)\cdot\tau(t)\right\}\right],\nonumber
\end{eqnarray}
\end{widetext}
where we in the above have used the following definitions
\begin{widetext}
\begin{eqnarray}
\int \{\rmd h\; \rmd \hat{h}\; \rmd g\; \rmd \hat{g}\}&=&\prod_{t=0}^{t_{\rm max}\!-\!1}\left\{\prod_{i=1}^{N^\sigma}\left\{\int\frac{\mathrm d  h_{i}(t)\;\mathrm d  \hat{h}_{i}(t)}{2\pi}\right\}\prod_{\ell=1}^{N^\tau}\left\{\int\frac{\mathrm d  g_{\ell}(t)\;\mathrm d  \hat{g}_{\ell}(t)}{2\pi}\right\}\right\},
\end{eqnarray}
\begin{eqnarray}
&&\Prob \big[\sigma(t_{\rm max}),\tau(t_{\rm max})\leftarrow\cdots\leftarrow\sigma(1),\tau(1)\vert h(t_{\rm max}\!-\!1),g(t_{\rm max}\!-\!1),\cdots,h(0),g(0)\big]\nonumber\\
&&=\prod_{t=0}^{t_{\rm max}\!-\!1}\left\{\prod_{i=1}^{N^\sigma}\frac{\rme^{\beta\sigma_i(t\!+\!1)h_i(t)}}{2\cosh[\beta h_i(t)]}\prod_{\ell=1}^{N^\tau}\frac{\rme^{\beta\tau_\ell(t\!+\!1)g_\ell(t)}}{2\cosh[\beta g_\ell(t)]}\right\}\label{def:P}
\end{eqnarray}
\end{widetext}
and, to reduce notation, we used various variables in a vector form ($h(t)=(h_1(t),\ldots,h_N(t))$, etc.) wherever possible.

In order to proceed with the computation of (\ref{eq:G1}), we have to specify the interaction variables  in the term
\begin{widetext}
\begin{eqnarray}
&&\exp\left[-\rmi\sum_{t=0}^{t_{\rm max}\!-\!1}\left\{\hat{h}(t)\cdot h(\bsigma(t),\btau(t))+\hat{g}(t)\cdot g(\bsigma(t),\btau(t))\right\}\right]\label{def:disor}\\
&&=\prod_{t=0}^{t_{\rm max}\!-\!1}\exp\left[-\rmi\sum_{i=1}^{N^{\sigma}}\hat{h}_i(t)\left\{\sum_{j\neq i}^{N^\sigma}J_{ij}^{\sigma}\sigma_j(t)+\sum_{j\neq i}^{N^\tau}J_{ij}^{\sigma\leftarrow\tau}\tau_j(t)+\theta_i^\sigma(t)\right\}\right]\nonumber\\
&&\times\exp\left[-\rmi\sum_{\ell=1}^{N^{\tau}}\hat{g}_\ell(t)\left\{  \sum_{k\neq \ell}^{N^\tau}J_{\ell k}^{\tau}\tau_k(t)+\sum_{k\neq \ell}^{N^\sigma}J_{\ell k}^{\tau\leftarrow\sigma}\sigma_k(t)+\theta_\ell^\tau(t)                    \right\}\right]\nonumber
\end{eqnarray}
\end{widetext}
which contains all information about the structure of our model. For the sake of simplicity we 
 take the interactions $J_{ij}^\sigma=\frac{J^\sigma}{N^\sigma}$, $J_{ij}^{\sigma\leftarrow\tau}=\frac{J^{\sigma\leftarrow\tau}}{N^\tau}$ and  $J_{ij}^\tau$,  $J_{ij}^{\tau\leftarrow\sigma}$ are random independent variables drawn from the distributions  $\frac{1}{2}\delta(J_{ij}^\tau-\frac{ J^\tau}{ \sqrt{N^\tau}})+\frac{1}{2}\delta(J_{ij}^\tau+\frac{ J^\tau}{ \sqrt{N^\tau}})$ and $\frac{1}{2}\delta(J_{ij}^{\tau\leftarrow\sigma}-\frac{ J^{\tau\leftarrow\sigma}}{ \sqrt{N^\sigma}})+\frac{1}{2}\delta(J_{ij}^{\tau\leftarrow\sigma}+\frac{ J^{\tau\leftarrow\sigma}}{ \sqrt{N^\sigma}})$ respectively.  The scaling of these interactions will allow us to take the thermodynamic limit later on.  First, however, we have to deal with the disorder in interactions. Assuming that the system is self-averaging~\cite{Dominicis} (which is expected for a very large system) allows us to take disorder averages in (\ref{def:disor}).

\subsection{Disorder averages\label{section:DA}}

Let us now take the averages in the disorder-dependent part of (\ref{def:disor})

\begin{widetext}
\begin{eqnarray}
&&\overline{\prod_{t=0}^{t_{\rm max}\!-\!1}\exp\left[-\rmi\sum_{\ell=1}^{N^{\tau}}\hat{g}_\ell(t)\left\{  \sum_{k\neq \ell}^{N^\tau}J_{\ell k}^{\tau}\tau_k(t)+\sum_{k\neq \ell}^{N^\sigma}J_{\ell k}^{\tau\leftarrow\sigma}\sigma_k(t)                   \right\}\right]}^{\{J_{\ell k}^{\tau},J_{\ell k}^{\tau\leftarrow\sigma}\}}\label{eq:disordAv}\\
&&=\overline{ \prod_{\ell=1}^{N^{\tau}}
\prod_{k\neq \ell}^{N^\tau}\exp\left[-\rmi J_{\ell k}^{\tau}\sum_{t=0}^{t_{\rm max}\!-\!1}\hat{g}_\ell(t)\tau_k(t)\right]}^{\{J_{\ell k}^{\tau}\}}\overline{\prod_{k\neq \ell}^{N^\sigma}\exp\left[-\rmi J_{\ell k}^{\tau\leftarrow\sigma}\sum_{t=0}^{t_{\rm max}\!-\!1}\hat{g}_\ell(t)\sigma_k(t)\right]}^{\{J_{\ell k}^{\tau\leftarrow\sigma}\}}\nonumber\\
&&=\prod_{\ell=1}^{N^{\tau}}
\prod_{k\neq \ell}^{N^\tau}\cos\left[\frac{ J^\tau}{ \sqrt{N^\tau}}\sum_{t=0}^{t_{\rm max}\!-\!1}\hat{g}_\ell(t)\tau_k(t)\right]
\prod_{k\neq \ell}^{N^\sigma}\cos\left[\frac{ J^{\tau\leftarrow\sigma}}{ \sqrt{N^\sigma}}\sum_{t=0}^{t_{\rm max}\!-\!1}\hat{g}_\ell(t)\sigma_k(t)\right]\nonumber\\
&&=\prod_{\ell=1}^{N^{\tau}}
\exp\left[\!-\!\frac{ 1}{ 2N^\tau}\sum_{k\neq \ell}^{N^\tau}\left(J^\tau\sum_{t=0}^{t_{\rm max}\!-\!1}\hat{g}_\ell(t)\tau_k(t)\right)^2\right]\exp\left[\!-\!\frac{ 1}{ 2N^\sigma}\sum_{j\neq \ell}^{N^\sigma}\left(J^{\tau\leftarrow\sigma}\sum_{t=0}^{t_{\rm max}\!-\!1}\hat{g}_\ell(t)\sigma_j(t)\right)^2\!+\!O(N^0)\right]\nonumber\\
&&=\prod_{\ell=1}^{N^{\tau}}
\exp\left[\!-\!\frac{ (J^\tau)^2}{ 2N^\tau}\sum_{k\neq \ell}^{N^\tau}\sum_{t, t^\prime}\hat{g}_\ell(t)\hat{g}_\ell(t^\prime)\tau_k(t)\tau_k(t^\prime)\right]\exp\left[\!-\!\frac{ (J^{\tau\leftarrow\sigma})^2}{ 2N^\sigma}\sum_{j\neq \ell}^{N^\sigma}\sum_{t, t^\prime}\hat{g}_\ell(t)\hat{g}_\ell(t^\prime)\sigma_j(t)\sigma_j(t^\prime)\!+\!O(N^0)\right]\nonumber
\end{eqnarray}
\end{widetext}
In the above we have used the asymptotic identity $\cos(x)=\exp(-\frac{x^2}{2}+ O(x^4))$ as $x\rightarrow0$ to obtain the quadratic form in the last line of (\ref{eq:disordAv}). We note that for any interactions of the form $\frac{J}{\sqrt{N}}$  with random $J$ sampled from the (well behaved) distribution $\Prob(J)$, with $\int\rmd J \;\Prob(J)J=0$ and $\int\rmd J\; \Prob(J)J^2=1$, the result of the disorder average (\ref{eq:disordAv}) remains the same.

\subsection{Order parameters\label{section:OP}}

Using the scaling of interactions from the section \ref{section:GF} and the results of disorder averages from the section \ref{section:DA} we obtain the disorder-averaged generating functional
\begin{widetext}
\begin{eqnarray}
\overline{\Gamma[\psi^\sigma,\psi^\tau]}&=&\int \{\rmd h\; \rmd \hat{h}\; \rmd g\; \rmd \hat{g}\}\exp\left[\rmi\sum_{t=0}^{t_{\rm max}\!-\!1}\left\{\hat{h}(t)\cdot\left[h(t)-\theta^\sigma(t)\right]+\hat{g}(t)\cdot\left[g(t)-\theta^\tau(t)\right]\right\}\right]\label{eq:G2}\\
&&\times\!\!\!\!\sum_{\{\sigma_i(t),\tau_i(t)\}}\!\!\!\! \exp\left[-\rmi\sum_{i=1}^{N^{\sigma}}\sum_{t=0}^{t_{\rm max}\!-\!1}\hat{h}_i(t)\left\{\frac{J^{\sigma}}{N^\sigma}\sum_{j\neq i}^{N^\sigma}\sigma_j(t)\!+\!\frac{J^{\sigma\leftarrow\tau}}{N^\tau}\sum_{k\neq i}^{N^\tau}\tau_k(t)\right\}\right]\nonumber\\
&&\times\prod_{\ell=1}^{N^{\tau}}\exp\left[\!-\!\frac{ (J^\tau)^2}{ 2N^\tau}\sum_{k\neq \ell}^{N^\tau}\sum_{t, t^\prime}\hat{g}_\ell(t)\hat{g}_\ell(t^\prime)\tau_k(t)\tau_k(t^\prime)\right]\exp\left[\!-\!\frac{ (J^{\tau\leftarrow\sigma})^2}{ 2N^\sigma}\sum_{j\neq \ell}^{N^\sigma}\sum_{t, t^\prime}\hat{g}_\ell(t)\hat{g}_\ell(t^\prime)\sigma_j(t)\sigma_j(t^\prime)\!+\!O(N^0)\right]\nonumber\\
&&\times\Prob \big[\sigma(t_{\rm max}),\tau(t_{\rm max})\leftarrow\cdots\leftarrow\sigma(1),\tau(1)\nonumber\\
&&\vert h(t_{\rm max}-1),g(t_{\rm max}-1),\cdots,h(0),g(0)\big]\Prob(\sigma(0))\Prob(\tau(0))\exp\left[-\rmi\sum_{t=0}^{t_{\rm max}}\left\{\psi^\sigma(t)\cdot\sigma(t)+\psi^\tau(t)\cdot\tau(t)\right\}\right]\nonumber
\end{eqnarray}
\end{widetext}
Inserting into above the following representations of unity for all times $t$
\begin{widetext}
 \begin{eqnarray}
&&\int\frac{\mathrm d  m^\sigma(t)\;\mathrm d  \hat{m}^\sigma(t)}{2\pi/N^\sigma}\;\rme^{\rmi  N^\sigma\hat{m}^\sigma(t)[ m^\sigma(t)-\frac{1}{N^\sigma}\sum_{i=1}^{N^\sigma}\sigma_i(t)]}=1\label{def:orderPar}\\
&&\int\frac{\mathrm d  m^\tau(t)\;\mathrm d  \hat{m}^\tau(t)}{2\pi/N^\tau}\;\rme^{\rmi  N^\tau\hat{m}^\tau(t)[ m^\tau(t)-\frac{1}{N^\tau}\sum_{i=1}^{N^\tau}\tau_i(t)]}=1\nonumber
\end{eqnarray}
\begin{eqnarray}
&&\int\frac{\mathrm d  q^\sigma(t,t^\prime)\;\mathrm d  \hat{q}^\sigma(t,t^\prime)}{2\pi/N^\sigma}\;\rme^{\rmi  N^\sigma\hat{q}^\sigma(t,t^\prime)[ q^\sigma(t,t^\prime)-\frac{1}{N^\sigma}\sum_{i=1}^{N^\sigma}\sigma_i(t)\sigma_i(t^\prime)]}=1\nonumber\\
&&\int\frac{\mathrm d  q^\tau(t,t^\prime)\;\mathrm d  \hat{q}^\tau(t,t^\prime)}{2\pi/N^\tau}\;\rme^{\rmi  N^\tau\hat{q}^\tau(t,t^\prime)[ q^\tau(t,t^\prime)-\frac{1}{N^\tau}\sum_{i=1}^{N^\tau}\tau_i(t)\tau_i(t^\prime)
]}\!=\!1\nonumber
\end{eqnarray}
\end{widetext}
and using that they are just integrals over the $\delta$-functions in their Fourier representation, we obtain
\begin{widetext}
\begin{eqnarray}
&&\overline{\Gamma[\psi^\sigma,\psi^\tau]}=\int\left\{    \mathrm d  m^\sigma     \mathrm d  \hat{m}^\sigma      \mathrm d  m^\tau    \mathrm d  \hat{m}^\tau      \mathrm d  q^\sigma   \mathrm d  \hat{q}^\sigma    \mathrm d  q^\tau    \mathrm d  \hat{q}^\tau\right\}\exp\left[ \rmi  N^\sigma\sum_{t}\hat{m}^\sigma(t)\; m^\sigma(t) +     \rmi  N^\sigma\sum_{t,t^\prime}\hat{q}^\sigma(t,t^\prime)\;  q^\sigma(t,t^\prime) \right]\label{eq:G3}\\
&&\times\exp\left[ \rmi  N^\tau\sum_{t}\hat{m}^\tau(t)\;  m^\tau(t) +  \rmi  N^\tau\sum_{t,t^\prime}\hat{q}^\tau(t,t^\prime)\; q^\tau(t,t^\prime)\right]\nonumber\\
&&\times\sum_{\{\sigma_i(t),\tau_i(t)\}}
\prod_{t=0}^{t_{\rm max}\!-\!1}\prod_{i=1}^{N^{\sigma}} \Biggl\{\int \rmd h_i(t)\;\delta\left(h_i(t)-J^{\sigma} m^\sigma(t)-J^{\sigma\leftarrow\tau} m^\tau(t)-\theta_i^\sigma(t)+\Delta_i^h(\sigma,\tau)\right)\Biggr\}\nonumber\\
%
&&\times\int \{ \rmd g\; \rmd \hat{g}\}\prod_{\ell=1}^{N^\tau}
\exp\left[\rmi\sum_{t=0}^{t_{\rm max}\!-\!1}\hat{g}_\ell(t)\left[g_\ell(t)-\theta_\ell^\tau(t)\right]\right]\exp\left[\!-\!\frac{ 1}{ 2}(J^\tau)^2\sum_{t, t^\prime}\hat{g}_\ell(t)\mathrm{A}(t,t^\prime)
\hat{g}_\ell(t^\prime)                 +\Delta_\ell^\mathrm{A} (\sigma,\tau)\right]\nonumber\\
&&\times\prod_{i=1}^{N^\sigma}\exp\left[  -\rmi \sum_t\hat{m}^\sigma(t) \sigma_i(t) - \rmi  \sum_{t,t^\prime}\hat{q}^\sigma(t,t^\prime) \sigma_i(t)\sigma_i(t^\prime)                     \right]\prod_{i=1}^{N^\tau}\exp\left[- \rmi  \sum_t\hat{m}^\tau(t) \tau_i(t) -  \rmi  \sum_{t,t^\prime}\hat{q}^\tau(t,t^\prime) \tau_i(t)\tau_i(t^\prime)             \right]\nonumber\\
&&\times\Prob \big[\sigma(t_{\rm max}),\tau(t_{\rm max})\leftarrow\cdots\leftarrow\sigma(1),\tau(1)\nonumber\\
&&~~~~~\vert h(t_{\rm max}-1),g(t_{\rm max}-1),\cdots,h(0),g(0)\big]\Prob(\sigma(0))\Prob(\tau(0))\exp\left[-\rmi\sum_{t=0}^{t_{\rm max}}\left\{\psi^\sigma(t)\cdot\sigma(t)+\psi^\tau(t)\cdot\tau(t)\right\}+O(N^0)\right],\nonumber
\end{eqnarray}
\end{widetext}
where in the above we have used the following notations
 \begin{eqnarray}
&&\int\left\{    \mathrm d  m^\sigma     \mathrm d  \hat{m}^\sigma      \mathrm d  m^\tau    \mathrm d  \hat{m}^\tau      \mathrm d  q^\sigma   \mathrm d  \hat{q}^\sigma    \mathrm d  q^\tau    \mathrm d  \hat{q}^\tau\right\}\equiv\\
&&\int\frac{\mathrm d  m^\sigma(t)\;\mathrm d  \hat{m}^\sigma(t)}{2\pi/N^\sigma}\int\frac{\mathrm d  m^\tau(t)\;\mathrm d  \hat{m}^\tau(t)}{2\pi/N^\tau}\nonumber\\
&&\times \int\frac{\mathrm d  q^\sigma(t,t^\prime)\;\mathrm d  \hat{q}^\sigma(t,t^\prime)}{2\pi/N^\sigma}\; \int\frac{\mathrm d  q^\tau(t,t^\prime)\;\mathrm d  \hat{q}^\tau(t,t^\prime)}{2\pi/N^\tau}\;\nonumber
\end{eqnarray}
and
\begin{eqnarray}
\mathrm{A}(t,t^\prime)=  q^\tau(t,t^\prime)       +\left[\frac{J^{\tau\leftarrow\sigma}}{J^\tau}\right]^2 q^\sigma(t,t^\prime) . \label{def:A}
\end{eqnarray}
The corrections $ \Delta_i^h(\sigma,\tau)=\frac{J^{\sigma}}{N^\sigma}\sigma_i(t) + \frac{J^{\sigma\leftarrow\tau}}{N^\tau}\tau_i(t)$ and 

\begin{widetext}
\[\Delta_\ell^\mathrm{A}(\sigma,\tau)=\frac{ 1}{ 2}(J^\tau)^2\sum_{t, t^\prime}\hat{g}_\ell(t)
\Biggl\{\frac{ 1}{ N^\tau}\tau_\ell(t)\tau_\ell(t^\prime)
+\frac{ 1}{ N^\sigma}\left[\frac{J^{\tau\leftarrow\sigma}}{J^\tau}\right]^2\sigma_\ell(t)\sigma_\ell(t^\prime) \Biggr\}
\hat{g}_\ell(t^\prime)\]
\end{widetext}
 contribute to the $O(N^0)$ term in the equation  (\ref{eq:G3}).
\newpage

Using the Gaussian integral identity
\begin{widetext}
\begin{eqnarray}
\exp\left[\!-\!\frac{ 1}{ 2}(J^\tau)^2\sum_{t, t^\prime}\hat{g}(t)\mathrm{A}(t,t^\prime)
\hat{g}(t^\prime)                  \right] 
=\sqrt{\frac{\vert A^{-1}\vert}{\left(2\pi\right)^{t_{\rm max}}}}\int\{\rmd\phi\}\exp\left[ -\frac{1}{2} \sum_{t,t^\prime} \phi(t)\mathrm{A}^{-1}(t,t^\prime)\phi(t^\prime)  -\rmi J^\tau\sum_{t}\phi(t)\hat{g}(t)\right] \label{def:Gauss} 
\end{eqnarray}
\end{widetext}
allows us to linearise the quadratic form in the equation (\ref{eq:G3}). This with subsequent integration over the $\hat{g}$ variables gives us
\begin{widetext}
\begin{eqnarray}
\overline{\Gamma[\psi^\sigma,\psi^\tau]}&=&\int\left\{    \rmd  m^\sigma     \mathrm d  \hat{m}^\sigma      \mathrm d  m^\tau    \mathrm d  \hat{m}^\tau      \mathrm d  q^\sigma   \mathrm d  \hat{q}^\sigma    \mathrm d  q^\tau    \mathrm d  \hat{q}^\tau\right\}\exp\left[ \rmi  N^\sigma\sum_{t}\hat{m}^\sigma(t)\; m^\sigma(t) +     \rmi  N^\sigma\sum_{t,t^\prime}\hat{q}^\sigma(t,t^\prime)\;  q^\sigma(t,t^\prime) \right]\label{eq:G4}\\
&&\times\exp\left[+ \rmi  N^\tau\sum_{t}\hat{m}^\tau(t)\;  m^\tau(t) +  \rmi  N^\tau\sum_{t,t^\prime}\hat{q}^\tau(t,t^\prime)\; q^\tau(t,t^\prime)\right]\nonumber\\
&&\times\!\!\!\!\sum_{\{\sigma_i(t),\tau_i(t)\}}\!\!\!\!
\prod_{t=0}^{t_{\rm max}\!-\!1}\Biggl\{\prod_{i=1}^{N^{\sigma}}\int \rmd h_i(t)\delta\left(h_i(t)\!-\!J^{\sigma} m^\sigma(t)\!-\!J^{\sigma\leftarrow\tau} m^\tau(t)\!-\!\theta_i^\sigma(t)\right)\Biggr\}\nonumber\\
&&\times\prod_{\ell=1}^{N^\tau}\Biggl\{
\sqrt{\frac{\vert A^{-1}\vert}{\left(2\pi\right)^{t_{\rm max}}}}\int\{\rmd\phi\}\exp\left[ -\frac{1}{2} \sum_{t,t^\prime} \phi(t)\mathrm{A}^{-1}(t,t^\prime)\phi(t^\prime)\right]\prod_{t=0}^{t_{\rm max}\!-\!1}\int\rmd g_\ell(t)\;\delta\left( g_\ell(t)-J^\tau\phi(t)-\theta_\ell^\tau(t)        \right)\Biggr\}\nonumber\\
&&\times\prod_{i=1}^{N^\sigma}\exp\left[  -\rmi \sum_t\hat{m}^\sigma(t) \sigma_i(t) - \rmi  \sum_{t,t^\prime}\hat{q}^\sigma(t,t^\prime) \sigma_i(t)\sigma_i(t^\prime)                     \right]\prod_{i=1}^{N^\tau}\exp\left[- \rmi  \sum_t\hat{m}^\tau(t) \tau_i(t) -  \rmi  \sum_{t,t^\prime}\hat{q}^\tau(t,t^\prime) \tau_i(t)\tau_i(t^\prime)             \right]\nonumber\\
&&\times\Prob \big[\sigma(t_{\rm max}),\tau(t_{\rm max})\leftarrow\cdots\leftarrow\sigma(1),\tau(1)\vert h(t_{\rm max}-1),g(t_{\rm max}-1),\cdots,h(0),g(0)\big]\Prob(\sigma(0))\Prob(\tau(0))\nonumber\\
&&\times\exp\left[-\rmi\sum_{t=0}^{t_{\rm max}}\left\{\psi^\sigma(t)\cdot\sigma(t)+\psi^\tau(t)\cdot\tau(t)\right\}+O(N^0)\right]\nonumber\\
&&=\int\left\{    \mathrm d  m^\sigma     \mathrm d  \hat{m}^\sigma      \mathrm d  m^\tau    \mathrm d  \hat{m}^\tau      \mathrm d  q^\sigma   \mathrm d  \hat{q}^\sigma    \mathrm d  q^\tau    \mathrm d  \hat{q}^\tau\right\}\exp\left[ \rmi  N^\sigma\sum_{t}\hat{m}^\sigma(t)\; m^\sigma(t) +     \rmi  N^\sigma\sum_{t,t^\prime}\hat{q}^\sigma(t,t^\prime)\;  q^\sigma(t,t^\prime) \right]\nonumber\\
&&\times\exp\left[ \rmi  N^\tau\sum_{t}\hat{m}^\tau(t)\;  m^\tau(t) +  \rmi  N^\tau\sum_{t,t^\prime}\hat{q}^\tau(t,t^\prime)\; q^\tau(t,t^\prime)\right]\nonumber\\
&&\times\sum_{\{\sigma_i(t),\tau_i(t)\}}\exp\left[O(N^0)\right]
\prod_{i=1}^{N^\sigma}\exp\left[  -\rmi \sum_t\hat{m}^\sigma(t) \sigma_i(t) - \rmi  \sum_{t,t^\prime}\hat{q}^\sigma(t,t^\prime) \sigma_i(t)\sigma_i(t^\prime)  \right]\exp\left[-\rmi\sum_{t=0}^{t_{\rm max}}\psi_i^\sigma(t)\;\sigma_i(t)\right]\nonumber\\
&&\times\prod_{i=1}^{N^\sigma}\left\{\prod_{t=0}^{t_{\rm max}\!-\!1}\frac{\rme^{\beta\sigma_i(t\!+\!1)\left\{J^{\sigma} m^\sigma(t)+J^{\sigma\leftarrow\tau} m^\tau(t)+\theta_i^\sigma(t)\right\}}}{2\cosh[\beta \left\{J^{\sigma} m^\sigma(t)+J^{\sigma\leftarrow\tau} m^\tau(t)+\theta_i^\sigma(t)\right\}]}\right\}\frac{1}{2}\left[1+m^\sigma(0)\;\sigma_i(0)\right]\nonumber\\
&&\times\prod_{\ell=1}^{N^\tau}\exp\left[- \rmi  \sum_t\hat{m}^\tau(t) \tau_\ell(t) -  \rmi  \sum_{t,t^\prime}\hat{q}^\tau(t,t^\prime) \tau_\ell(t)\tau_\ell(t^\prime)             \right]\exp\left[-\rmi\sum_{t=0}^{t_{\rm max}}\psi_\ell^\tau(t)\;\tau_\ell(t)\right]\nonumber\\
&&\times\prod_{\ell=1}^{N^\tau}\sqrt{\frac{\vert A^{-1}\vert}{\left(2\pi\right)^{t_{\rm max}}}}\int\{\rmd\phi\}\exp\left[ -\frac{1}{2} \sum_{t,t^\prime} \phi(t)\mathrm{A}^{-1}(t,t^\prime)\phi(t^\prime)\right]\nonumber\\
&&\times\left\{\prod_{t=0}^{t_{\rm max}\!-\!1}\frac{\rme^{\beta\tau_\ell(t\!+\!1)\left\{J^\tau\phi(t)+\theta_\ell^\tau(t)\right\}}}{2\cosh[\beta\left\{ J^\tau\phi(t)+\theta_\ell^\tau(t)\right\}]}\right\}\frac{1}{2}\left[1+m^\tau(0)\;\tau_\ell(0)\right]\nonumber
\end{eqnarray}
\end{widetext}
\begin{widetext}
Let us now define the two objects
\begin{eqnarray}
&&\mathrm{M}^\sigma\left[\{\sigma_i(t)\}\vert\left\{\psi_i^\sigma(t)\right\}\right]=\exp\left[  -\rmi \sum_t\hat{m}^\sigma(t)\; \sigma_i(t) - \rmi  \sum_{t,t^\prime}\hat{q}^\sigma(t,t^\prime)\; \sigma_i(t)\;\sigma_i(t^\prime)  \right]
\exp\left[-\rmi\sum_{t=0}^{t_{\rm max}}\psi_i^\sigma(t)\;\sigma_i(t)\right]\label{def:Msigma}\\
&&\times\left\{\prod_{t=0}^{t_{\rm max}\!-\!1}\frac{\rme^{\beta\sigma_i(t\!+\!1)\left\{J^{\sigma} m^\sigma(t)+J^{\sigma\leftarrow\tau} m^\tau(t)+\theta_i^\sigma(t)\right\}}}{2\cosh[\beta \left\{J^{\sigma} m^\sigma(t)+J^{\sigma\leftarrow\tau} m^\tau(t)+\theta_i^\sigma(t)\right\}]}\right\}\frac{1}{2}\left[1+m^\sigma(0)\;\sigma_i(0)\right]\nonumber\\\nonumber\\
&&\mathrm{M}^\tau\left[\{\tau_\ell(t)\}\vert\left\{\psi_\ell^\tau(t)\right\}\right]=\exp\left[- \rmi  \sum_t\hat{m}^\tau(t)\; \tau_\ell(t) -  \rmi  \sum_{t,t^\prime}\hat{q}^\tau(t,t^\prime)\; \tau_\ell(t)\;\tau_\ell(t^\prime)             \right]\exp\left[-\rmi\sum_{t=0}^{t_{\rm max}}\psi_\ell^\tau(t)\;\tau_\ell(t)\right]\label{def:Mtau}\\
&&\times\sqrt{\frac{\vert A^{-1}\vert}{\left(2\pi\right)^{t_{\rm max}}}}\int\{\rmd\phi\}\exp\left[ -\frac{1}{2} \sum_{t,t^\prime} \phi(t)\mathrm{A}^{-1}(t,t^\prime)\phi(t^\prime)\right]\left\{\prod_{t=0}^{t_{\rm max}\!-\!1}\frac{\rme^{\beta\tau_\ell(t\!+\!1)\left\{J^\tau\phi(t)+\theta_\ell^\tau(t)\right\}}}{2\cosh[\beta \left\{J^\tau\phi(t)+\theta_\ell^\tau(t)\right\}]}\right\}\frac{1}{2}\left[1+m^\tau(0)\;\tau_\ell(0)\right].\nonumber
\end{eqnarray}
\end{widetext}
Using above definitions in the final result of  (\ref{eq:G4}), with $N^\sigma=\gamma N$ and $N^\tau=(1-\gamma)N$ we are able to write the disorder-averaged generating functional (\ref{eq:G4}) in the form of an integral
\begin{widetext}
\begin{eqnarray}
&&\overline{\Gamma[\psi^\sigma,\psi^\tau]}=\int\left\{    \mathrm d  m^\sigma     \mathrm d  \hat{m}^\sigma      \mathrm d  m^\tau    \mathrm d  \hat{m}^\tau      \mathrm d  q^\sigma   \mathrm d  \hat{q}^\sigma    \mathrm d  q^\tau    \mathrm d  \hat{q}^\tau\right\}\rme^{N\Psi[m^\sigma, \hat{m}^\sigma,  m^\tau, \hat{m}^\tau,   q^\sigma,  \hat{q}^\sigma ,   q^\tau,\hat{q}^\tau, \psi^\sigma,\psi^\tau]+O(N^0)},\label{eq:Integral}
\end{eqnarray}
where 
\begin{eqnarray}
&&\Psi[m^\sigma, \hat{m}^\sigma,  m^\tau, \hat{m}^\tau,   q^\sigma,  \hat{q}^\sigma ,   q^\tau,\hat{q}^\tau, \psi^\sigma,\psi^\tau]\label{def:psi}\\
&&= \rmi  \gamma\sum_{t}\hat{m}^\sigma(t)\; m^\sigma(t) +     \rmi  \gamma\sum_{t,t^\prime}\hat{q}^\sigma(t,t^\prime)\;  q^\sigma(t,t^\prime) + \rmi  (1-\gamma)\sum_{t}\hat{m}^\tau(t)\;  m^\tau(t) +  \rmi  (1-\gamma)\sum_{t,t^\prime}\hat{q}^\tau(t,t^\prime)\; q^\tau(t,t^\prime)\nonumber\\
&&+\frac{1}{N}\sum_{i=1}^{N^\sigma}\log\left[\sum_{\{\sigma_i(t)\}}\mathrm{M}^\sigma\left[\{\sigma_i(t)\}\vert\left\{\psi_i^\sigma(t)\right\}\right]\right]+\frac{1}{N}\sum_{i=1}^{N^\tau}\log\left[\sum_{\{\tau_i(t)\}}\mathrm{M}^\tau\left[\{\tau_i(t)\}\vert\left\{\psi_i^\tau(t)\right\}\right]\right].\nonumber
\nonumber
\end{eqnarray}
\end{widetext}
Now for $N\rightarrow\infty$ we can use the saddle-point method to evaluate this integral.

\subsection{Saddle-point problem}

The integral (\ref{eq:Integral}) is dominated by the extrema of the function (\ref{def:psi}). To obtain  these  we solve the equations $\frac{\partial\Psi}{\partial \Omega}=0$, where $\Omega\in\{m^\sigma, \hat{m}^\sigma,  m^\tau, \hat{m}^\tau,   q^\sigma,  \hat{q}^\sigma ,   q^\tau,\hat{q}^\tau\}$, which gives us
\begin{eqnarray}
&&m^\sigma(t)=\left\langle \sigma(t)\right\rangle_{\M^\sigma}\label{eq:sp}\\
&&\rmi\hat{m}^\sigma(t)=\beta J^\sigma\Big(\left\langle \sigma(t\!+\!1)\right\rangle_{\M^\sigma}\nonumber\\
&&~~~~~~~~~-\tanh\left[\beta   \left\{J^{\sigma} m^\sigma(t)+J^{\sigma\leftarrow\tau} m^\tau(t)+\theta^\sigma(t)\right\}          \right]\Big)\nonumber\\
&&m^\tau(t)=\left\langle \tau(t)\right\rangle_{\M^\tau}\nonumber\\
&&\rmi\hat{m}^\tau(t)=\frac{\gamma}{1-\gamma}\beta J^{\sigma\leftarrow\tau}\Big(\left\langle \sigma(t\!+\!1)\right\rangle_{\M^\sigma}\nonumber\\
&&~~~~~-\tanh\left[\beta   \left\{J^{\sigma} m^\sigma(t)+J^{\sigma\leftarrow\tau} m^\tau(t)\right\}          \right]\Big)\nonumber\\
&&q^\sigma(t,t^\prime)=\left\langle \sigma(t)\;\sigma(t^\prime)\right\rangle_{\M^\sigma}\nonumber\\
&&q^\tau(t,t^\prime)=\left\langle \tau(t)\;\tau(t^\prime)\right\rangle_{\M^\tau}\nonumber\\
&&\rmi\hat{q}^\tau(t,t^\prime)=\rmi\hat{q}^\sigma(t,t^\prime)=0\nonumber,
\end{eqnarray}
where the averages $\left\langle \cdots\right\rangle_{\M^\sigma}$ and $\left\langle \cdots\right\rangle_{\M^\tau}$ are generated by the (site independent)  weight functions\footnote{We have removed (set to zero) generating fields in the functional (\ref{def:psi}), external fields in the $\tau$-system and set $\theta_i^\sigma(t)=\theta^\sigma(t)$ in the $\sigma$-system.} (\ref{def:Msigma}) and (\ref{def:Mtau}) respectively.


In order to show that the equality $\rmi\hat{q}^\tau(t,t^\prime)=\rmi\hat{q}^\sigma(t,t^\prime)=0$ is true, we first, using the Gaussian identity (\ref{def:Gauss}),  rewrite the equation (\ref{def:Mtau}) as follows
\begin{widetext}
\begin{eqnarray}
\mathrm{M}^\tau\left[\{\tau_\ell(t)\}\vert\left\{\psi_\ell^\tau(t)\right\}\right]&=&\int \{ \rmd g_\ell(t)\; \rmd \hat{g}_\ell(t)\}\;\mathrm{M}^\tau\left[\{\tau_\ell(t)\};\{g_\ell(t),\hat{g}_\ell(t)\}\vert\left\{\psi_\ell^\tau(t)\right\}\right]\label{def:Mtau2ndDef}\\
&&=\int \{ \rmd g_\ell(t)\; \rmd \hat{g}_\ell(t)\}\exp\left[- \rmi  \sum_t\hat{m}^\tau(t)\; \tau_\ell(t) -  \rmi  \sum_{t,t^\prime}\hat{q}^\tau(t,t^\prime)\; \tau_\ell(t)\;\tau_\ell(t^\prime)             \right]\nonumber\\
&&
\times\exp\left[\rmi\sum_{t=0}^{t_{\rm max}\!-\!1}\hat{g}_\ell(t)\left[g_\ell(t)-\theta_\ell^\tau(t)\right]\right]\exp\left[\!-\!\frac{ 1}{ 2}(J^\tau)^2\sum_{t, t^\prime}\hat{g}_\ell(t)\mathrm{A}(t,t^\prime)
\hat{g}_\ell(t^\prime)                  \right]\nonumber\\
&&\times\left\{\prod_{t=0}^{t_{\rm max}\!-\!1}\frac{\rme^{\beta\tau_\ell(t\!+\!1)g_\ell(t)}}{2\cosh[\beta g_\ell(t)]}\right\}\frac{1}{2}\left[1+m^\tau(0)\;\tau_\ell(0)\right]\exp\left[-\rmi\sum_{t=0}^{t_{\rm max}}\psi_\ell^\tau(t)\;\tau_\ell(t)\right],\nonumber
\end{eqnarray}
\end{widetext}
where $\mathrm{A}(t,t^\prime)=  q^\tau(t,t^\prime)    +\left[\frac{J^{\tau\leftarrow\sigma}}{J^\tau}\right]^2 q^\sigma(t,t^\prime) $.
\begin{widetext}
From the above it is clear that
\begin{eqnarray}
\frac{\partial}{\partial q(t,t^\prime)}\log\left[\sum_{\{\tau_\ell(t)\}} \int \{ \rmd g_\ell(t)\; \rmd \hat{g}_\ell(t)\}\;\M^\tau\left[\ldots\right] \right]=\!-\!\frac{ 1}{ 2}J^2\left\langle\hat{g}_\ell(t)
\hat{g}_\ell(t^\prime)\right\rangle_{\M^\tau}, \label{eq:Dq}
\end{eqnarray}
but also
\begin{eqnarray}
\!-\!\frac{\partial^2}{\partial \theta_\ell^\tau(t)\partial \theta_\ell^\tau(t^\prime)}\log\left[\sum_{\{\tau_\ell(t)\}} \int \{ \rmd g_\ell(t)\; \rmd \hat{g}_\ell(t)\}\;\M^\tau\left[\ldots\right] \right]=\left\langle\hat{g}_\ell(t)
\hat{g}_\ell(t^\prime)\right\rangle_{\M^\tau} .\label{eq:Dtheta}
\end{eqnarray}
\end{widetext}
Now using the above results and the identity $\frac{\partial^2}{\partial \theta_\ell^\tau(t)\partial \theta_\ell^\tau(t^\prime)}\Gamma[0,0]=0$  (since $\Gamma[0,0]=1$),  it is not difficult to show that the equality $\rmi\hat{q}^\tau(t,t^\prime)=\rmi\hat{q}^\sigma(t,t^\prime)=0$ is true. Application of this equality in the equations (\ref{eq:sp}) leads to further simplifications after which we obtain the following four equations
\begin{widetext}
\begin{eqnarray}
m^\sigma(t+1)&=& \tanh\left[\beta   \left\{J^{\sigma} m^\sigma(t)+J^{\sigma\leftarrow\tau} m^\tau(t)+\theta^\sigma(t)\right\}\right]\label{eq:m_sigma}\\
m^\tau(t+1)&=&\sqrt{\frac{\vert A^{-1}\vert}{\left(2\pi\right)^{t_{\rm max}}}}\int\{\rmd\phi\}\exp\left[ -\frac{1}{2} \sum_{t,t^\prime} \phi(t)\mathrm{A}^{-1}(t,t^\prime)\phi(t^\prime)\right]\tanh\left[\beta J^\tau\phi(t)\right]\label{eq:m_tau}\\
q^\sigma(t,t^\prime)&=&\delta_{t,t^\prime}+(1-\delta_{t,t^\prime})\;m^\sigma(t)\;m^\sigma(t^\prime)\label{eq:q_sigma}\\
q^\tau(t+1,t^\prime+1)&=&\sqrt{\frac{\vert A^{-1}\vert}{\left(2\pi\right)^{t_{\rm max}}}}\int\{\rmd\phi\}\exp\left[ -\frac{1}{2} \sum_{t,t^\prime} \phi(t)\mathrm{A}^{-1}(t,t^\prime)\phi(t^\prime)\right]\tanh\left[\beta J^\tau\phi(t)\right]\tanh\left[\beta J^\tau\phi(t^\prime)\right]\label{eq:q_tau}
\end{eqnarray}
\end{widetext}
Now the multivariate Gaussian probability measure in the equation (\ref{eq:m_tau}) can be reduced to a Gaussian of one variable only (with zero mean) thereby leading us to the result  ($3$).

\section{Dynamics of a sparse model: Cayley tree\label{section:sparse}}
\begin{figure}
\begin{center}
\setlength{\unitlength}{1.4mm}
\begin{picture}(60,55)
\put(0,0){\epsfysize=45\unitlength\epsfbox{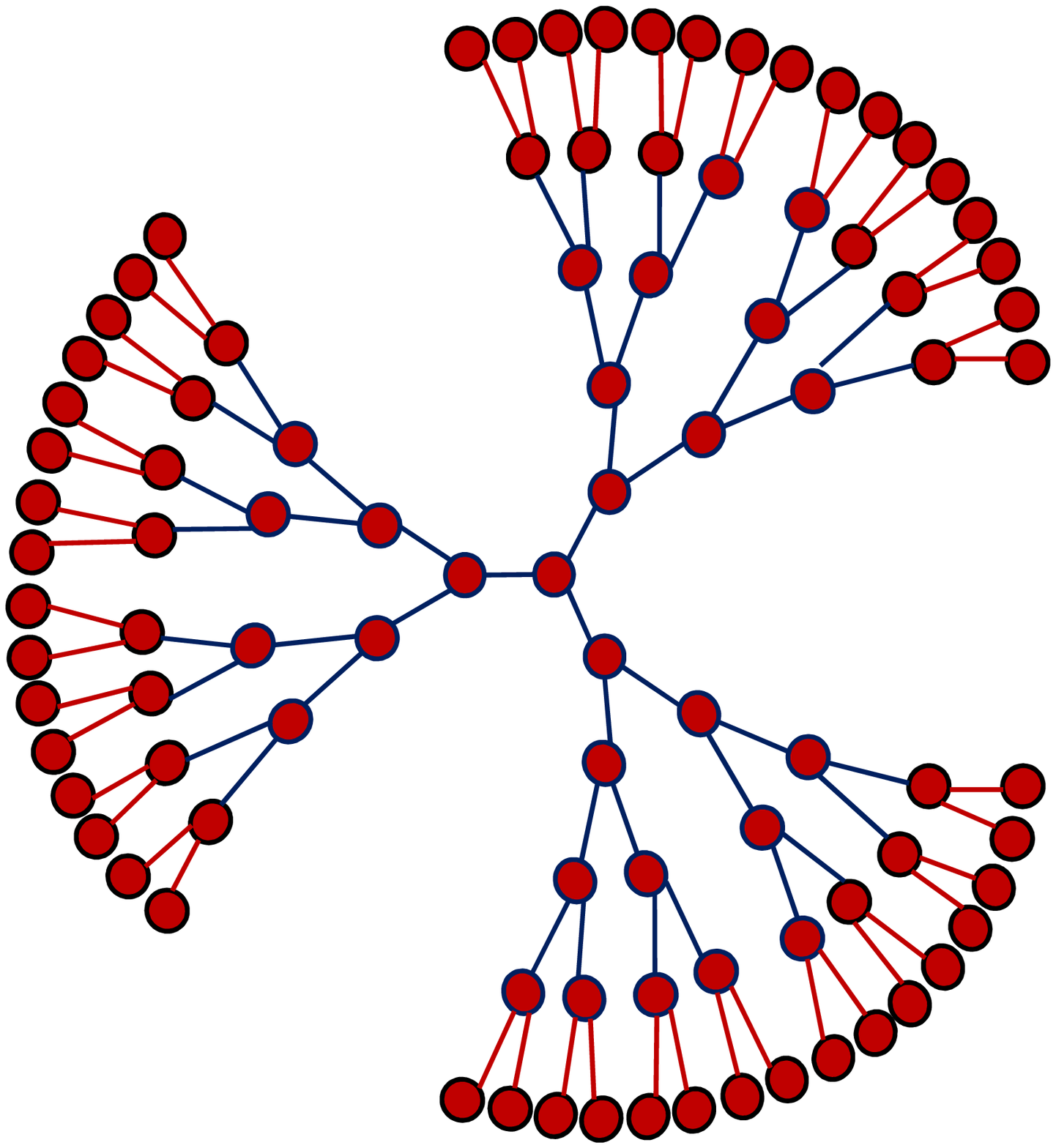}}
\end{picture}
\end{center}
\caption{Cayley tree with $k=3$ and $r=4$ generations (edges painted in blue) with the asymmetric boundary (edges painted in red).\label{fig:3}}
\end{figure}

In this section we study the dynamics of  Ising spin system which is governed by the Markov process of Eq.~(\ref{eq:master}). This system has a  Cayley tree topology of degree $k$ with $r$ generations; all edges in this tree are symmetric except the boundary edges which are asymmetric (see Figure~\ref{fig:3}). The sites on the boundary are subject to the random time-dependent fields  $\theta_i(t)\in\{-1,1\}$, where $\Prob(\theta_i(t)=\pm1)=1/2$.

Without the contribution of asymmetric boundary the process (\ref{eq:master}) is converging to the equilibrium Gibbs-Boltzmann distribution with Hamiltonian~(\ref{def:Hg-b}). For the ferromagnetic system with $J_{ij}=J>0$ and $\theta_i=0$ the free energy per spin $f(\beta)=-\frac{1}{\beta}\log 2 \cosh(\beta J)$ (that gives the average internal energy per spin $\left\langle E\right\rangle=\frac{\partial}{\partial\beta}\beta f(\beta)=-\tanh(\beta J)$) is an analytic function of the inverse temperature $\beta=\frac{1}{T}$, which rules out a phase transition in this system for any $T>0$~\cite{Eggarter}. 

\begin{figure}
%
\setlength{\unitlength}{1.4mm}
\begin{picture}(60,55)
\put(0,0){\epsfysize=45\unitlength\epsfbox{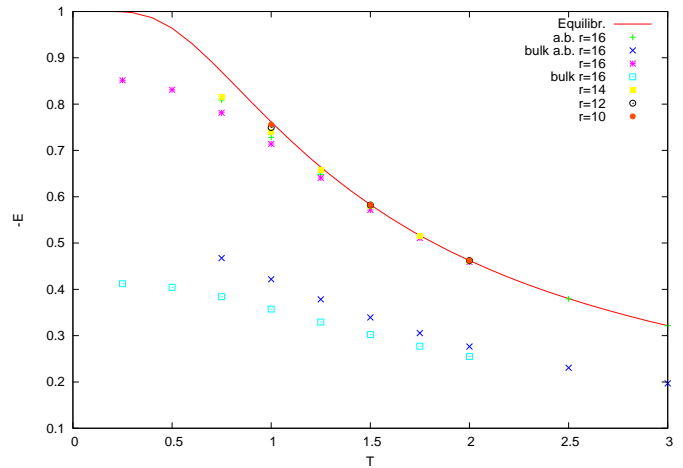}}
\end{picture}
\caption{Comparing the equilibrium energy $E=-\tanh(J/T)$ (solid line) with the energy measured ($E(\sigma)=-\frac{1}{N}\sum_{\langle ij\rangle}J_{ij}\sigma_i
\sigma_j$) in Monte Carlo simulation (symbols) on the symmetric part of a Cayley tree (of degree $k=3$ and of radius $r=19$) with asymmetric boundary. For this system $T_c=0$. The measurements are taken away from the asymmetric boundary (with incoming edges, pointing towards the center, denoted a.b. in the figure) on a sub-tree of radius $r=\{10,12,14, 16\}$. For comparison, the value of  $E(\sigma)$ for the total system (symbols labeled by `bulk') is included. The case of a boundary with incoming edges is also compared with that of a  boundary with equal (on average) number of incoming  and outgoing edges.
\label{fig:4}}
\end{figure}

Adding an asymmetric boundary disturbs the detailed balance condition and there is no guarantee that the system will end up in a thermal equilibrium state  asymptotically.  Nevertheless, the symmetric part of the system (provided that it is at a sufficient distance from the asymmetric boundary)  exhibits equilibrium like behavior as can be seen in  Figure~\ref{fig:4}.
%

Results obtained in the Figure~\ref{fig:4} are also valid if a Cayley tree is embedded in the following random graph topology. Suppose we generate a very large random regular graph of degree $k$ ($N$ being the number of nodes). The number of short loops (of a finite length)   is vanishing with increasing $N$ and only long loops of order $O(\log N)$ are present in this graph~\cite{circuits}. By following the neighbors of an arbitrary node in this network and its neighbors of neighbors, etc.  we can form a Cayley tree of radius $r$.

Suppose we pick one of these Cayley trees and make all the edges belonging to it symmetric and the rest of the edges in the network asymmetric (incoming with probability $1/2$). The dynamics of the Ising spin on a Cayley tree is dominated by the dynamics of its boundary which is described by the set $\{m_i(t)\}$ of local magnetizations $m_i(t)=\sum_\sigma \Prob_t(\sigma)\;\sigma_i$, which in a very large system ($N\rightarrow\infty$ with $r=O(N^0)$) are dominated by the asymmetric part of this system. However, we have shown in the Letter that after long time these local magnetizations are vanishing  and the results obtained for the original Cayley tree configuration holds also here.



\end{document}